\documentclass[12pt,twoside]{elsart}
\usepackage{fleqn}

\usepackage{graphicx}
\usepackage{amssymb}

\usepackage[figuresright]{rotating}

\newcommand{\AmS}{{\protect\the\textfont2
  A\kern-.1667em\lower.5ex\hbox{M}\kern-.125emS}}

\begin{document}
\begin{frontmatter}

\title{Renormalisation in Quantum Mechanics, Quantum Instantons and Quantum Chaos}

\author[MCSD]{H. Jirari }
\author[MCSD] {H. Kr\"oger\thanksref{mailHK} }
\author[MCSDS] { X.Q. Luo }
\author[Mor] { K.J.M. Moriarty }

\address[MCSD]{D\'epartement de Physique, Universit\'e Laval, Qu\'ebec, Qu\'ebec G1K 7P4, Canada}%
\address[MCSDS]{Department of Physics, Zhongshan University, Guangzhou 510275, China}%
\address[Mor] {Department of Mathematics, Statistics and Computer Science, \\
        Dalhousie University, Halifax, Nova Scotia B3H 3J5, Canada}
\thanks[mailHK] {invited talk given by H. Kr\"oger. Email: hkroger@phy.ulaval.ca.}


\begin{abstract}
We suggest how to construct non-perturbatively a renormalized action in 
quantum mechanics. We discuss similarties and differences with the standard 
effective action. We propose that the new quantum action is suitable to 
define and compute quantum instantons and quantum chaos.
\end{abstract}
\end{frontmatter}

\section{INTRODUCTION}
Quantum mechanics, which describes physics at atomic length scales can not be understood by the laws of classical physics valid at macroscopic length scales.
Examples are: Heisenberg's uncertainty principle, quantum tunneling, Schr\"odinger's cat paradox, entangled states, Einstein-Rosen-Podolski paradox, quantum cryptology, quantum computing etc. On the other hand, in modern physics there are notions which have proven to be quite useful and which have their origin in classical physics.
Examples are instantons and chaos. Instantons play a role in quantum chromodynamics (QCD), the standard model of strong interactions. They may be important for the mecanism of confinement of quarks. Presumably they play an important role in nuclear matter at high temperature and density, where a phase transition from the hadronic phase to the quark-gluon plasma has been predicted. Even a richer phase structure may exist \cite{Shuryak}. Furthermore, in the inflationary scenario of the early universe, instantons are important. For a review see Ref.\cite{Linde}. 
During inflation, quantum fluctuations of the primordial field expand exponentially and eventually end up as a classical field. The fluctuations are of the size of the horizon \cite{Starobinsky}. The classical fluctuations eventually lead to galaxy formation \cite{Khlopov}.
Chaos has been observed by a huge number of phenomena in macroscopic i.e. classical physics. But chaotic phenomena were also found in systems ruled by quantum mechanics. For an overview see Ref. \cite{Blumel}.
Examples are: The hydrogen atom in a strong magnetic field, 
showing strong irregularities in its spectrum \cite{Friedrich}.
Irregular patterns have been found in the wave functions of the quantum mechanical model of the stadium billard \cite{McDonald}.
Billard like boundary conditions have been realized experimentally 
in mesoscopic quantum systems, like quantum dots and quantum corrals, formed by atoms in semi-conductors \cite{Stockmann}.
So what is the problem with instantons and chaos in quantum physics?
It has to do with its proper definition. The underlying reason is due to the dynamical group of time evolution. In classical mechanics time evolution of a system can be viewed as an infinite sequence of infinitesimal canonical transformations. The corresponding dynamical group is the symplectic group. In quantum mechanics, a system governed by a time independent Hamiltonian, follows the time evolution of the unitary group. This difference has simple but drastic consequences: In classical physics, chaos is characterized, e.g. by Lyapunov exponents or Poincar\'e sections. This is based on identifying trajectories in phase space (position and conjugate momentum).
In quantum mechanics, Heisenberg's uncertainty relation $\Delta x \Delta p \ge 
\hbar/2$ does not allow to specify a point in phase space with zero error!
Consequently, the apparatus of classical chaos theory can not be simply taken over to quantum physics. The same problem occurs also with instantons.
Let us consider a 1-D system in quantum mechanics with a particle of mass $m$ moving in a potential $V(x)=A(x^{2}- a^{2})^{2}$. This potential has two minima at $x=\pm a$. The instanton $x_{inst}(t)$ is the solution of the classical equation of motion in imaginary time, with boundary conditions such that the particle starts at $x(t=-\infty)=-a$, $\dot{x}(t=-\infty)=0$ and arrives at $x(t=+\infty)=+a$, $\dot{x}(t=+\infty)=0$. The problem again is that quantum mechanics does not allow to specify both, position and momentum with zero uncertainty. 
Due to this problem, workers in quantum chaos have tried to characterize such systems in different ways, alternative to those of classical chaos. One successful route has been to characterize the spectral density of quantum system with chaotic classical counterpart by Poisson versus Wigner distributions. There is a conjecture by Bohigas et al. \cite{Bohigas}, which says that the signature of a classical chaotic system is a the spectral density following a Wigner distribution.

\subsection{Bridge between classical and quantum physics}
A different way to look at the problem of quantum chaos is the following. 
Quantum mechanics is a branch of physics which can be looked upon 
from many view points. There is the historic duality between the particle and wave interpretation (de Broglie). But there is a more modern view point: Renormalisation and the effective action. In short terms it means that the action of a system in quantum physics can be written as the action of the classical system, however, with modified parameters (mass, potential parameters).  

\begin{figure}[htb]
\label{fig:Weak.Anh}
\setlength{\textwidth}{6.5in}
\setlength{\textheight}{9.25in}
\setlength{\leftmargin}{0.25in}
\setlength{\oddsidemargin}{0.2in}
\setlength{\evensidemargin}{0.2in}
\setlength{\headheight}{0in}
\setlength{\headsep}{0in}
\setlength{\footskip}{0.25in}
\setlength{\parindent}{3em}
\setlength{\topmargin}{-0.25in}
\vspace*{\fill}
\begin{center}
\begingroup\makeatletter\ifx\SetFigFont\undefined%
\gdef\SetFigFont#1#2#3{%
\reset@font\fontsize{#1}{#2pt}%
\fontfamily{#3}
\selectfont}%
\fi\endgroup%
\resizebox{!}{15cm}{
\begin{picture}(0,0)%
\includegraphics{Weak.Anh.pstex}
\end{picture}%
\setlength{\unitlength}{0.24pt}%
\begingroup\makeatletter\ifx\SetFigFont\undefined%
\gdef\SetFigFont#1#2#3{%
\reset@font\fontsize{#1}{#2pt}%
\fontfamily{#3}
\selectfont}%
\fi\endgroup%
\begin{picture}(2173,2632)(126,334)
\put(1200,2800){\large{$\blacktriangledown\quad$ Numerical result}}
\put(1200,2700){\large{$\ast\quad$ Pertubation theory}}
\put(1200,2600){\large{T$=$4.0}}
\put(5,2600){\Huge{$\tilde{V}_4$}}
\put(1200,1950){\large{$\blacktriangledown\quad$ Numerical result}}
\put(1200,1850){\large{$\ast\quad$ Pertubation theory}}
\put(1200,1750){\large{T$=$4.0}}
\put(5,1700){\Huge{$\tilde{V}_2$}}
\put(1200,900){\large{T$=$4.0}}
\put(5,800){\Huge{$\tilde{m}$}}
\put(1200,200){\Huge{$\lambda$}}
\put(200,100){\LARGE{Figure 1. Harmonic oscillator with weak anharmonic perturbation.}}
\end{picture}
}
\end{center}
\vskip 20 mm
\vspace*{\fill}
\end{figure}

Renormalisation and the effective action is useful to describe an equivalent low-energy theory, starting from a high-energy theory. The construction of a low-energy theory in nuclear physics has been discussed by Lepage \cite{Lepage}.
The effective action in connection with the linear sigma model is presented 
in  Ref.\cite{Donoghue}. 
In a more narrow sense the effective action $\Gamma$
has been introduced in quantum field theory in such a way gives an expectation value $<\phi>$ which minimizes the potential energy, giving the ground state energy \cite{Jona,Coleman}. 

\begin{table}[htb]
\caption{Renormalized parameters of quartic potential.}
\label{tab:Quartic}
\begin{tabular}{|l|l|l|l|l|l|l|} \hline \hline
$\tilde{m}$ & $\tilde{v_{0}}$ & $\tilde{v_{1}}$ & $\tilde{v_{2}}$ & $\tilde{v_{3}}$ & $\tilde{v_{4}}$ & \mbox{interval} \\ \hline
0.9936(3) & 1.171(2) & 0.000(7) & 0.449(18) & 0.000(16) & 0.982(23) &  
-1.0,+1.0 \\ \hline 
0.9938(2) & 1.166(2) & 0.000(7) & 0.488(15) & 0.000(13) & 0.954(15) &  
-1.2,+1.2 \\ \hline 
0.9941(2) & 1.170(2) & 0.000(7) & 0.466(12) & 0.000(11) & 0.975(10) &  
-1.4,+1.4 \\ \hline 
0.9940(2) & 1.170(2) & 0.000(8) & 0.469(10) & 0.000(10) & 0.973(7) &  
-1.6,+1.6 \\ \hline 
0.9944(2) & 1.168(2) & 0.000(8) & 0.458(9) & 0.000(9) & 0.984(5) &  
-1.8,+1.8 \\ \hline 
0.9942(2) & 1.168(3) & 0.000(9) & 0.459(9) & 0.000(9) & 0.986(4) &  
-2.0,+2.0 \\ \hline 
0.9939(2) & 1.172(3) & 0.000(9) & 0.449(8) & 0.000(8) & 0.992(4) &  
-2.2,+2.2 \\ \hline 
0.9938(2) & 1.169(3) & 0.000(10) & 0.460(9) & 0.000(8) & 0.987(3) &  
-2.4,+2.4 \\ \hline 
0.9942(3) & 1.166(3) & 0.000(11) & 0.447(9) & 0.000(8) & 0.993(3) &  
-2.6,+2.6 \\ \hline
0.9938(3) & 1.174(3) & 0.000(12) & 0.432(10) & 0.000(8) & 0.998(3) &  
-2.8,+2.8 \\ \hline 
0.9946(3) & 1.171(3) & 0.000(13) & 0.435(10) & 0.000(9) & 0.990(3) &  
-3.0,+3.0 \\ \hline 
0.9940(4) & 1.170(3) & 0.000(13) & 0.461(11) & 0.000(9) & 0.984(3) &  
-3.2,+3.2 \\ \hline 
0.9943(4) & 1.165(4) & 0.000(15) & 0.460(11) & 0.000(10) & 0.985(4) &  
-3.4,+3.4 \\ \hline 
0.9949(5) & 1.161(4) & 0.000(16) & 0.483(12) & 0.000(11) & 0.973(4) &  
-3.6,+3.6 \\ \hline \hline
0.9941(3) & 1.169(3) & 0.000(10) & 0.458(11) & 0.000(10) & 0.983(7) & Mean \\ \hline \hline  
\end{tabular} \\[2pt]
{Transition time $T=0.5$, $J=6$ initial and final boundary points.}
\end{table}

An effective action has been also considered at finite temperature \cite{Dolan}.
Because the effective action has a mathematical structure similar to the classical action, and the quantum effects are taken into account by 
parameters different from their classical counter parts, the effective action looks like the ideal way to bridge the gap from quantum to classical physics
and eventually solve the quantum chaos and quantum instanton problem.
However, there is a catch. The effective potential and the effective action 
in quantum mechanics has been computed using perturbation theory by  
Cametti et al. \cite{Cametti}. Consider the Lagrangian
\begin{eqnarray}
L(q,\dot{q},t) &=& \frac{m}{2} \dot{q}^{2} - V(q), ~~~ 
V(q) = \frac{m}{2} \omega^{2} q^{2} + U(q) ,
\end{eqnarray}
and $U(q)$ is, say, a quartic potential $U(q) \sim q^{4}$.
Then the effective action is obtained in doing a loop $(\hbar)$ expansion
\begin{eqnarray}
\Gamma[q] &=& \int dt \left( - V^{eff}(q(t)) \right.
\nonumber \\
&+& \left. \frac{Z(q(t))}{2} \dot{q}^{2}(t)
+ A(q(t)) \dot{q}^{4}(t) + B(q(t)) (d^{2}q/dt^{2})^{2}(t) + \cdots \right)
\nonumber \\
V^{eff} &=& \frac{1}{2}m \omega^{2} q^{2} +U(q) + \hbar V^{eff}_{1}(q) + O(\hbar^{2})
\nonumber \\
Z(q) &=& m + \hbar Z_{1}(q) + O(\hbar^{2})
\nonumber \\
A(q) &=& \hbar A_{1}(q) + O(\hbar^{2})
\nonumber \\
B(q) &=& \hbar B_{1}(q) + O(\hbar^{2}) .
\end{eqnarray}
There are higher loop corrections to the effective potential $V^{eff}$ as well as to the mass renormalisation $Z$. The most important property is the 
occurrence of higher time derivative terms. Actually, there is an infinite series of increasing order. Here comes the problem. When we want to interpret $\Gamma$ as effective action, the higher time derivatives require
more intial/boundary conditions than the classical action. 
This is a catastrophy.
In the following we will present an alternative way to construct an action 
taking into acount quantum corrections.

\begin{figure}[htb]
\label{fig:Quartic}
\setlength{\textwidth}{6.5in}
\setlength{\textheight}{9.25in}
\setlength{\leftmargin}{0.25in}
\setlength{\oddsidemargin}{0.2in}
\setlength{\evensidemargin}{0.2in}
\setlength{\headheight}{0in}
\setlength{\headsep}{0in}
\setlength{\footskip}{0.25in}
\setlength{\parindent}{3em}
\setlength{\topmargin}{-0.25in}
\vspace*{\fill}
\begin{center}
\begingroup\makeatletter\ifx\SetFigFont\undefined%
\gdef\SetFigFont#1#2#3{%
\reset@font\fontsize{#1}{#2pt}%
\fontfamily{#3}
\selectfont}%
\fi\endgroup%
\resizebox{!}{15cm}{
\begin{picture}(0,0)%
\includegraphics{Quartic.pstex}
\end{picture}%
\setlength{\unitlength}{0.24pt}%
\begingroup\makeatletter\ifx\SetFigFont\undefined%
\gdef\SetFigFont#1#2#3{%
\reset@font\fontsize{#1}{#2pt}%
\fontfamily{#3}
\selectfont}%
\fi\endgroup%
\begin{picture}(2065,2752)(207,215)
\put(1700,2300){\large{$\circ\quad\tilde{V}_4$ M.C.}}
\put(1700,2200){\large{$\triangle\quad\tilde{V}_4$ S.E. 30 states}}
\put(1700,2100){\large{$\lozenge\quad\tilde{V}_4$ S.E. 7 states}}
\put(1100,2300){\large{$\bullet\quad\tilde{V}_2$ M.C.}}
\put(1100,2200){\large{$\blacktriangle\quad\tilde{V}_2$ S.E. 30 states}}
\put(1100,2100){\large{$\blacklozenge\quad\tilde{V}_2$ S.E. 7 states}}
\put(1700,1400){\large{$\circ\quad\tilde{V}_0$ M.C.}}
\put(1700,1300){\large{$\triangle\quad\tilde{V}_0$ S.E. 30 states}}
\put(1700,1200){\large{$\diamond\quad\tilde{V}_0$ S.E. 7 states}}
\put(1100,1400){\large{$\bullet\quad\tilde{m}$ M.C.}} 
\put(1100,1300){\large{$\blacktriangle\quad\tilde{m}$ S.E. 30 states}}
\put(1100,1200){\large{$\blacklozenge\quad\tilde{m}$ S.E. 7 states}} 
\put(1350,1000){\large{$----\quad E_0$}}
\put(300,100) {\LARGE{Figure 2. Renormalized parameters of quartic potential.}}
\end{picture}
}
\end{center}
\vskip 20 mm
\vspace*{\fill}
\end{figure}

\section{THE QUANTUM ACTION}
We suggest to construct a renormalized or quantum action from transition matrix elements, which involve the time evolution \cite{Jirari:a}. 
In Q.M. the transition amplitude from 
$x_{in}$, $t_{in}$ to $x_{fi}$, $t_{fi}$ is given by 
\begin{equation}
\label{PathIntegral}
G(x_{fi},t_{fi};x_{in},t_{in}) =
\left. \int [dx] \exp[ \frac{i}{\hbar} 
S[x] ] \right|_{x_{in},t_{in}}^{x_{fi},t_{fi}}  ,
\end{equation}
where 
\begin{equation}
S = \int dt  \frac{m}{2} \dot{x}^{2} - V(x) .
\end{equation}
%

\begin{table}[htb]
\caption{Renormalized parameters of double well potential.}
\label{tab:Double.Well}
\begin{tabular}{|l|l|l|l|l|l|l|l|} \hline \hline
$\tilde{m}$ & $\tilde{v_{0}}$ & $\tilde{v_{1}}$ & $\tilde{v_{2}}$ & $\tilde{v_{3}}$ & $\tilde{v_{4}}$ & \mbox{interval} \\ \hline
0.9959(1)  & 1.5701(54) & 0.000(2)  & -0.739(5)  & 0.000(2)  & 0.487(4)  &  -1.2,+1.2 \\ \hline
0.9961(2)  & 1.5714(17) & 0.000(2)  & -0.747(10) & 0.000(2)  & 0.489(7)  &  -1.4,+1.4 \\ \hline
0.9961(1)  & 1.5732(11) & 0.000(3)  & -0.760(5)  & 0.000(3)  & 0.499(3)  &  -1.6,+1.6 \\ \hline
0.9959(1)  & 1.5713(10) & 0.000(2)  & -0.747(4)  & 0.000(2)  & 0.495(2)  &  -1.8,+1.8 \\ \hline
0.9959(1)  & 1.5744(11) & 0.000(3)  & -0.754(4)  & 0.000(2)  & 0.498(2)  &  -2.0,+2.0 \\ \hline
0.9959(2)  & 1.5694(19) & 0.000(2)  & -0.739(6)  & 0.000(2)  & 0.492(3)  &  -2.2,+2.2 \\ \hline
0.9964(2)  & 1.5718(16) & 0.000(3)  & -0.745(6)  & 0.000(2)  & 0.491(2)  &  -2.4,+2.4 \\ \hline
0.9962(8)  & 1.5685(18) & 0.000(2)  & -0.740(7)  & 0.000(1)  & 0.492(3)  &  -2.6,+2.6 \\ \hline
0.9963(2)  & 1.5731(7)  & 0.000(0)  & -0.742(2)  & 0.000(2)  & 0.492(1)  &  -2.8,+2.8 \\ \hline
0.9966(3)  & 1.5695(2)  & 0.000(4)  & -0.744(8)  & 0.000(3)  & 0.492(2)) &  -3.0,+3.0 \\ \hline
0.9961(2) & 1.5710(17) & 0.000(2) & -0.745(6) & 0.000(2) & 0.493(3) & Mean \\ \hline \hline
\end{tabular} \\[2pt]
Transition time $T= 0.5$. $J=6$ initial and final boundary points.
\end{table}

\noindent denotes the classical action.
In some cases, this path integral can be expressed as a sum over 
classical paths only. 
\begin{equation}
\label{SumClassPath}
G(x_{fi},t_{fi}; x_{in},t_{in}) = 
\sum_{ \{x_{cl}\} } Z \exp \left[ \frac{i}{\hbar} 
\left. S[{x}_{cl}] \right|_{x_{in},t_{in}}^{x_{fi},t_{fi}} \right] .
\end{equation}
An example is the harmonic oscillator ($V(x) = \frac{m}{2} \omega^{2} x^{2}$),
where
\begin{eqnarray}
\left. S[x_{cl}] \right|_{x_{in},t_{in}}^{x_{fi},t_{fi}} 
&=& \frac{ m \omega}{2 \sin(\omega T) } 
\left[ (x_{fi}^{2} + x_{in}^{2}) \cos(\omega T) - 2 x_{in}x_{fi} \right] , 
\nonumber \\
Z &=&  \sqrt{ \frac { m \omega }{ 2 \pi i \hbar \sin(\omega T) } }, ~~~
T = t_{fi}-t_{in} , 
\end{eqnarray}
where $S[x_{cl}]$ denotes the classical action evaluated along the 
classical trajectory from $x_{in}$, $t_{in}$ to $x_{fi}$, $t_{fi}$. 
In the following we suggest to introduce the quantum action 
defined by a generalisation of Eq.(\ref{SumClassPath}) 
where the action has mathematically a structure like the classical action, 
and where quantum effects are taken into account by modified action parameters
(renormalisation).

{\it Conjecture}:
For a given classical action $S = \int dt \frac{m}{2} \dot{x}^{2} - V(x)$ 
there is a quantum action 
$\tilde{S} = \int dt \frac{\tilde{m}}{2} \dot{x}^{2} - \tilde{V}(x)$, 
which allows to express the transition amplitude by
\begin{equation}
\label{DefRenormAction}
G(x_{fi},t_{fi}; x_{in},t_{in}) = \tilde{Z} 
\exp [ \frac{i}{\hbar} \left. \tilde{S}[\tilde{x}_{cl}] 
\right|_{x_{in},t_{in}}^{x_{fi},t_{fi}} ] .
\end{equation}
Here $\tilde{x}_ {cl}$ denotes the classical path, such that the action $\tilde{S}(\tilde{x}_{cl})$ 
is minimal (we exclude the occurrence of conjugate points or caustics). 
$\tilde{Z}$ denotes the normalisation factor corresponding to $\tilde{S}$. 
Eq.(\ref{DefRenormAction}) is valid with 
the {\em same} action $\tilde{S}$ for all sets of 
boundary positions $x_{fi}$, $x_{in}$ for a given time interval $T=t_{fi}-t_{in}$. 
The parameters of the quantum action depend on the time $T$. The 
quantum action converges to a non-trivial limit when $T \to \infty$. 
Any dependence on $x_{fi}, x_{in}$ enters via the trajectory 
$\tilde{x}_ {cl}$. $\tilde{Z}$ depends on the action parameters and $T$, 
but not on $x_{fi}, x_{in}$.
\begin{figure}[htb]
\label{fig:Double.Well}
\setlength{\textwidth}{6.5in}
\setlength{\textheight}{9.25in}
\setlength{\leftmargin}{0.25in}
\setlength{\oddsidemargin}{0.2in}
\setlength{\evensidemargin}{0.2in}
\setlength{\headheight}{0in}
\setlength{\headsep}{0in}
\setlength{\footskip}{0.25in}
\setlength{\parindent}{3em}
\setlength{\topmargin}{-0.25in}
\vspace*{\fill}
\begin{center}
\begingroup\makeatletter\ifx\SetFigFont\undefined%
\gdef\SetFigFont#1#2#3{%
\reset@font\fontsize{#1}{#2pt}%
\fontfamily{#3}
\selectfont}%
\fi\endgroup%
\resizebox{!}{15cm}{
\begin{picture}(0,0)%
\includegraphics{Double_well.pstex}
\end{picture}%
\setlength{\unitlength}{0.24pt}%
\begingroup\makeatletter\ifx\SetFigFont\undefined%
\gdef\SetFigFont#1#2#3{%
\reset@font\fontsize{#1}{#2pt}%
\fontfamily{#3}
\selectfont}%
\fi\endgroup%
\begin{picture}(2129,2752)(161,215)
\put(1900,2800){\large{$\vartriangle\quad\tilde{V}_4$ M.C.}}
\put(1900,2700){\large{$\vartriangleright\quad\tilde{V}_4$ S.E.}}
\put(1500,2800){\large{$\blacktriangle\quad\tilde{V}_2$ M.C.}}
\put(1500,2700){\large{$\blacktriangleright\quad\tilde{V}_2$ S.E.}}
\put(1900,1400){\large{$\circ\quad\tilde{V}_0$ M.C.}}
\put(1900,1300){\large{$\lozenge\quad\tilde{V}_0$ S.E.}}
\put(1500,1400){\large{$\bullet\quad\tilde{m}$ M.C.}}
\put(1500,1300){\large{$\blacklozenge\quad\tilde{m}$ S.E.}}
\put(1700,1200){\large{$----\quad E_0$}}
\put(250,100){\LARGE{ Figure 3. Renormalized parameters of double well potential.}} 
\end{picture}
}
\end{center}
\vskip 20 mm
\vspace*{\fill}
\end{figure}
One may ask: what is the difference between effective and quantum action?
Conceptually, effective action and 
quantum action are quite similar.
However, its technical definition is different and also its physical content.
While the effective action corresponds to infinite time 
and allows to obtain the ground state energy, the quantum action 
is defined for arbitrary finite time $T$. In Euclidean formulation, the inverse time corresponds to temperature. Thus the quantum action allows to describe quantum physics at finite temperature 
(including excited states). However, the effective action can be defined also at finite temperature \cite{Dolan}.
The effective action can be computed analytically by perturbation theory (loop expansion). However, this series is not convergent. 
Practically, it can be used only for some small number of loops and small values of the perturbation parameter.
The quantum action can be computed non-perturbatively for all values of the coupling parameter. 
The effective action has the defect of generating higher order time derivatives.
The quantum action is postulated to be free of higher time derivative terms. 
To construct the quantum action being sensitive to excited states, one needs transition matrix elements beyond the vacuum sector. We have chosen to use position states in Q.M. In Q.F.T. this corresponds to Bargman states. The definition of the quantum action is related to the Schr\"odinger functional method in Q.F.T. \cite{SchrodFunct}.

\subsection{Quantum action at finite temperature}
First we make a Wick rotation to imaginary time. 
The purpose is, first to make the path integral well defined (Wiener measure)
allowing to apply Monte Carlo methods for its numerical computation. Secondly, the instanton is defined in imaginary time.
One effect of this transformation is that it changes a relative sign between 
the kinetic term and the potential term of the action. 
Thus in the following we work with imaginary time (Euclidean) actions and 
Green's functions.
Let us see how the quantum action is related to finite temperature physics. 
According to the laws of quantum mechanics and thermodymical equilibrium,
the expectation value of some observable $O$, like e.g. average energy
is given by
\begin{eqnarray}
<O> &=& \frac{ Tr\left[ O ~ \exp[ - \beta H] \right] }
{ Tr\left[ \exp[ - \beta H] \right] }
\nonumber \\
&=& \frac{ \int_{-\infty}^{+\infty} dx \int_{-\infty}^{+\infty} dy
<x|O|y><y|\exp[ - \beta H ]|x> }
{ \int_{-\infty}^{+\infty} dx <x|\exp[ - \beta H ]|x> } ,
\end{eqnarray}
where $\beta$ is related to the temperature $\tau$ by 
$\beta = 1/(k_{B} {\tau})$.
On the other hand the (Euclidean) transition amplitude is given by
\begin{equation}
G(x_{fi},T;x_{in},0) = <x_{fi}| \exp[ - H T/\hbar ]|x_{in}>
\end{equation}
Thus from the definition of the quantum action, Eq.(\ref{DefRenormAction}),
one obtains
\begin{equation}
<O> = \frac{ \int_{-\infty}^{+\infty} dx \int_{-\infty}^{+\infty} dy
<x|O|y> \exp[-\tilde{S}_{\beta}|_{x,0}^{y,\beta}] }
{ \int_{-\infty}^{+\infty} dx \exp[ - \tilde{S}_{\beta}|_{x,0}^{x,\beta}] } ,
\end{equation}
if we identify $\beta = \frac{1}{k_{B} {\tau}} = T/\hbar$.
As a result, the quantum action $\tilde{S}_{\beta}$ computed from transition time $T$, 
describes equilibrium thermodynamics at $\beta = T/\hbar$, i.e. temperature $\tau = 1/(k_{B} \beta)$.
As an important consequence, once having determined the quantum action for some finite time, one can use it for the study of quantum instantons and quantum chaos at finite temperature (see below).

\subsection{Construction of quantum action}
Suppose the classical action is given by
\begin{equation}
S = \int_{0}^{T} dt \frac{m}{2} \dot{x}^{2} + v_{4} x^{4}(t).
\end{equation}
Then we make an ansatz for the quantum action
\begin{equation}
\tilde{S} = \int_{0}^{T} dt \frac{\tilde{m}}{2} \dot{x}^{2} 
+ \tilde{v}_0 + \tilde{v}_{1} x(t) + \cdots + v_{N} x^{N}(t) .
\end{equation}
Then $\tilde{m}$, $\tilde{v}_0$, \dots, $\tilde{v}_{N}$ are the renormalized parameters which take into account the quantum corrections. 
Their values have been determined by making a global best fit to a number of transition amplitudes $G(x_j,T;x_i,0)$ (which satisfies Eq.\ref{DefRenormAction}), where $x_i$, $x_j$ haven been taken from a set of points $\{x_1,\cdots,x_J\}$ and those points have been chosen to cover
some interval $[-a,+a]$.  
More details are given in Ref.\cite{Jirari:a}.

\section{RESULTS}
(i) Firstly we have considered a harmonic oscillator with a weak anharmonic perturbation,
\begin{equation}
S = \int dt \frac{m}{2} \dot{x}^{2} + v_2 x^{2} + \lambda v_4 x^{4},
~~~ \hbar = m = v_2 = \frac{1}{2} m \omega^{2} = v_4 = 1.
\end{equation}
In the ansatz of the quantum action, we have (for sake of numerical effort)
restricted the parameter space of the potential to 4th order polynomials.
We have varied the parameter $\lambda$ in the range 
$0 \leq \lambda \leq 0.1$ (weak perturbation).
We have computed numerically the quantum action (at transition time $T=4$) 
and compared it with the predictions of low order perturbation theory of the effective action. The latter predicts to one loop order and up to 2nd order of the coupling parameter $\lambda$
\begin{eqnarray}
v^{eff}_2 &=&  v_2 + \delta v_2, ~~~ 
\delta v_2 = \frac{3}{m \omega} \lambda 
\nonumber \\
v^{eff}_4 &=& v_4 + \delta v_4, ~~~ 
\delta v_4 = \frac{9}{m \omega^{2}} \lambda^{2} .
\end{eqnarray}
The results are shown in Fig.\ref{fig:Weak.Anh}.
One observes that the effective potential and the potential of the quantum action are close for small $\lambda$. The discrepancy for $\lambda \approx 1$
is due to neglect of of higher order perturbative corrections, i.e. the onset of non perturbative effects. Moreover, one observes that for $T=0$, the quantum action agrees with the classical action. Odd terms $\tilde{v}_1$ and $\tilde{v}_3$ are compatible with zero, due to parity conservation.
(ii) Secondly, we have considered the quartic potential ($m = \hbar =1$) 
given by $V(x) = v_4 x^{4}$, $v_4 = 1$.
The parameters of the renormalized action are shown in 
Tab.\ref{tab:Quartic} and in Fig.2.
In Tab.\ref{tab:Quartic} the parameters are shown for transition time $T=0.5$
corresponding to temperature $\tau =2$ ($\hbar = k_{B} = 1$).
The table also shows the estimated error of the parameters from the fit and the value of $\chi^{2}$ of the fit. Also variation of parameters under variation of the boundary points (i.e. the interval $[-a,+a]$) is shown. The magnitude of variance is comparable to the magnitude of errors.  
Fig.2 shows the behavior of renormalized parameters as function of the transition time $T$ (inverse temperature).
The error bars represent the error estimates of the fit. One observes
that for $T \to 0$, the renormalized action parameters tend to their classical counter part (except for the $\tilde{v}_4$ term, which is presumably due to a numerical shortcoming). For $T \to \infty$, the renormalized parameters tend to some asymptotic values. The asymptotic regime is reached at about $T \approx 4.5$.
(iii) Finally, we have considered a double well potential, of the form
$V(x) = A (x^{2} -a^{2})^{2}$, with $a = 1$, $A = 1/2$. 
The parameters are equivalent to $v_0 = v_4 =1/2$, $v_2 = -1$.
Tab.\ref{tab:Double.Well} shows the renormalized parameters for $T=0.5$ under variation of the interval of initial and end points. 
Like in Tab.\ref{tab:Quartic} one observes the magnitude of the estimated errors compatible with the variance under variation of the size of the interval.
The quantum action parameters as a function of $T$ are displayed in Fig.3.
Like in Fig.2 one observes for $T \to 0$ that the renormalized parameters tend to their classical counterparts and for large $T$ ($T \approx 9$) there is a regime of asymptotic convergence.
The error bars in the figure are composed of two parts. The upper part represents the error estimated from the fit, while the lower part represents the $\sigma$ value of the estimator of the parameters under variation of the interval size. 
From both tables and Fig.3 we can see that the parameters of the quantum action is reasonnably independent of the initial and final points.
This was postulated in the conjecture. Hence the numerical results support the conjecture. For further discussion and more details see Ref.\cite{Jirari:a}.

\section{IMPLICATIONS ON INSTANTONS AND CHAOS}
As can be seen from Fig.3, for all $T$ in the range $0 \leq T \leq 9$ one observes that the quantum potential has also a double well structure like the classical potential
($\tilde{v}_0 > 0$, $\tilde{v}_4 > 0$, but $\tilde{v}_2 < 0$). 
Thus in all of this range it has an instanton solution. 
According to the Feynman-Kac formula, for $T \to \infty$ (corresponding to 
temperature $\tau \to 0$) the propagator is 
dominated by the ground state wave function and energy.
Thus the quantum instanton corresponding to temperature zero 
is approximately given by the instanton solution of the quantum potential with $T$ large ($T \approx 9$). Quantum instantons of larger temperature are those solutions corresponding to smaller values of $T$.
Such quantum instanton solutions have been computed in Ref.\cite{Jirari:b}.
As far as quantum chaos is concerned, 1-dimensional conservative systems with time-independent Hamiltonian do not generate classical chaos.
Thus they are not appropriate to search for quantum chaos.
However, the quantum action can be constructed for 2-dim Hamiltonian systems,
which are known to display classical chaos, like the Henon Heiles system, Paul-trap, etc. Thus we suggest to apply the tools of classical chaos theory (like Lyapunov exponents, Poincar\'e sections) to the quantum action of those systems\cite{Jirari:b}.

\noindent {\bf Acknowledgements} \\ 
H.K. and K.M. are grateful for support by NSERC Canada. 
X.Q.L. has been supported by NSF for Distinguished Young Scientists of China, by Guangdong Provincial NSF and by the Ministry of Education of China.

\end{document}